\documentclass[twocolumn,english,aps,prb]{revtex4}
\usepackage[T1]{fontenc}
\usepackage[latin9]{inputenc}
\usepackage{graphicx}
\usepackage{amssymb}

\makeatletter


\makeatother

\usepackage{babel}

\begin{document}

\title{Transfer of optical spectral weight in magnetically ordered superconductors}

\author{Rafael M. Fernandes}

\affiliation{Department of Physics and Astronomy and Ames Laboratory, Iowa State
University, Ames, Iowa 50011, USA }

\author{J\"org Schmalian}

\affiliation{Department of Physics and Astronomy and Ames Laboratory, Iowa State
University, Ames, Iowa 50011, USA }

\date{\today}

\begin{abstract}
We show that, in antiferromagnetic superconductors, the optical spectral
weight transferred to low frequencies below the superconducting transition
temperature originates from energies that can be much larger than
twice the superconducting gap $\Delta$. This contrasts to non-magnetic
superconductors, where the optical spectrum is suppressed only for
frequencies below $2\Delta$. In particular, we demonstrate that the
superfluid condensate of the magnetically ordered superconductor is
not only due to states of the magnetically reconstructed Fermi surface,
but is enhanced by transfer of spectral weight from the mid infrared
peak generated by the spin density wave gap. We apply our results
to the iron arsenide superconductors, addressing the decrease of the
zero-temperature superfluid density in the doping regime where magnetism
coexists with unconventional superconductivity. 
\end{abstract}


\maketitle

\section{Introduction}

Optical measurements reveal crucial spectral and electromagnetic properties
of a superconductor\cite{Mattis58,Ferrell58,Tinkham59}. For example,
the London penetration depth $\lambda$ can be determined from the
low frequency dependence of the imaginary part $\sigma^{\prime\prime}\left(\omega\right)$
of the optical conductivity $\sigma\left(\omega\right)$: \begin{equation}
\ \sigma^{\prime\prime}\left(\omega\rightarrow0\right)=\frac{c^{2}}{4\pi\lambda^{2}\omega},\label{Imsig}\end{equation}
where $c$ is the speed of light. In addition, the superconducting
gap $\Delta$ can be obtained from the real part of the optical conductivity,
$\sigma^{\prime}\left(\omega\right)$, since spectral weight is transferred
from energies below $2\Delta$ to the $\omega\rightarrow0$ contribution
$c^{2}\lambda^{-2}\delta\left(\omega\right)/4$. Kramers-Kronig transformation
of this $\delta$-function term then yields Eq.(\ref{Imsig}). These
effects can be illustrated if one starts from a Drude conductivity
\begin{equation}
\sigma^{\prime}\left(\omega\right)=\frac{\omega_{p}^{2}}{4\pi}\frac{\tau}{1+\left(\omega\tau\right)^{2}}\label{Drude}\end{equation}
in the normal state, with plasma frequency $\omega_{p}^{2}=4\pi e^{2}n/m^{\ast}$
and scattering time $\tau$. Here, $m^{\ast}$ is the optical mass
and $n$ is the electron density. In the superconducting state, the
entire weight of the Drude conductivity is transferred to the $\delta$-function
if $\Delta$ is larger than $\tau^{-1}$, leading to the result of
the BCS theory $\lambda^{-2}=\omega_{p}^{2}/c^{2}=4\pi e^{2}n/\left(m^{\ast}c^{2}\right)$
for clean superconductors\cite{BCS}. In the dirty limit, $\Delta\tau\ll1$,
the transfer of the spectral weight below $\omega=2\Delta$ can be
approximated by $2\Delta\times\sigma^{\prime}\left(0\right)$, yielding
the well known result for dirty superconductors\cite{AGD} $\lambda^{-2}=\frac{4}{\pi}\Delta\tau\omega_{p}^{2}/c^{2}$.
Consequently, the superfluid condensate \begin{equation}
n_{s}=\frac{m^{\ast}c^{2}}{4\pi e^{2}}\lambda^{-2}\ \ \end{equation}
is reduced compared to the particle density, $n_{s}/n=4\Delta\tau/\pi$.
The determination of $\Delta$ from the optical spectrum is most efficient
for $\Delta\tau\lesssim1$. Furthermore, the investigation of the
optical conductivity reveals crucial information in unconventional
superconductors . In the cuprate superconductors, the $f$-sum rule
\begin{equation}
\frac{\omega_{p}^{2}}{4}=\int_{-\infty}^{\infty}\sigma^{\prime}\left(\omega\right)d\omega\label{fsum}\end{equation}
was used to analyze whether the anomalous redistribution of spectral
weight below the superconducting transition temperature $T_{c}$ reveals
information about the change of the kinetic energy, or more precisely
of the optical mass $m^{\ast}$, upon entering the superconducting
state\cite{Basov05,Norman}. Finally, fine structures in the optical
spectrum were used to determine the mechanism of superconductivity
in the cuprates\cite{Schachinger,Abanov,Abanov2,vanderMarel}.

In the recently discovered FeAs superconductors\cite{Kamihara08,Rotter08},
the interplay of collective magnetic degrees of freedom and superconductivity
has attracted great interest, in particular given the strong evidence
for an electronic pairing mechanism with $s^{+-}$-pairing state\cite{Mazin08}.
In this state, the superconducting order parameter has opposite signs
in different sheets of the Fermi surface separated by the magnetic
ordering vector $\mathbf{Q}$. In distinction, in the conventional
$s^{++}$-pairing state, that is expected to originate from electron-phonon
coupling, the superconducting order parameter has the same sign everywhere.
The recent observations of the magnetic resonance mode\cite{Christianson08},
of the microscopic coexistence between magnetic and superconducting
order\cite{Fernandes10}, and, in particular, of the integer and
half-integer flux-quantum transitions in a niobium-iron pnictide loop\cite{Chen10}
give strong evidence for $s^{+-}$-pairing.

Important insights about the magnetic, superconducting and normal
states have also been obtained in measurements of $\sigma\left(\omega\right)$\cite{Li08,Hu08,Pfuner09,Wu10,Uchida10,Gorshunov10,Heuman10,Kim10,Lucarelli10}.
An analysis based on Eq.(\ref{Imsig}) and on the Ferrell-Glover-Tinkham
(FGT) sum rule\cite{Ferrell58,Tinkham59} (see Eq.(\ref{FGT}) below)
led to results for the penetration depth\cite{Li08,Uchida10} that are consistent
with the values obtained by other techniques (see, for example Ref.\cite{Gordon10}).
For instance, in $\mathrm{Ba}_{0.6}\mathrm{K}_{0.4}\mathrm{Fe}_{2}\mathrm{As}_{2}$,
which has $T_{c}=37\mathrm{K}$, the authors of Ref.\cite{Li08}
found $\lambda\simeq2000\mathrm{\mathring{A}}$ at $T=10\mathrm{K.}$
A typical value for the largest superconducting gap in the same compound
was estimated\cite{Li08} as $\Delta$ $\simeq10\mathrm{meV}$. Other
investigations resolved the individual gaps on the various Fermi surface
sheets \cite{Gorshunov10,Heuman10,Kim10}.

\begin{figure}

\begin{centering}
\includegraphics[width=1\columnwidth]{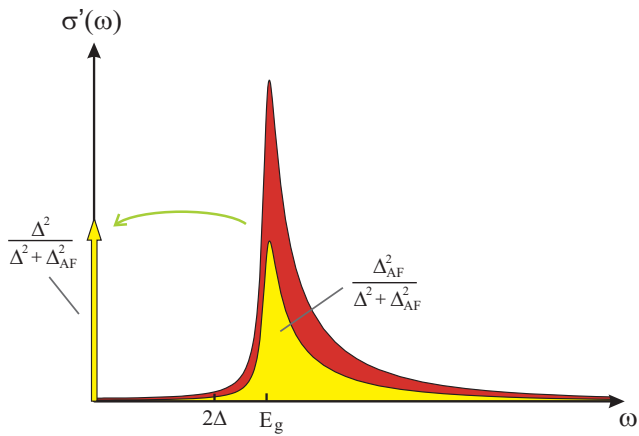} 
\par\end{centering}

\caption{Transfer of optical spectral weight for a magnetic superconductor,
considering particle-hole symmetry and a finite scattering rate $\tau^{-1}\simeq0.07\Delta_{\mathrm{AF}}$.
In the pure antiferromagnetic state (red), the spectrum has a peak
at $E_{g}=2\Delta_{\mathrm{AF}}$ and no Drude peak. In the coexistence
state (yellow), a fraction $\ \Delta^{2}/\left(\Delta_{\mathrm{AF}}^{2}+\Delta^{2}\right)$
of the total spectral weight of the antiferromagnetic phase is transferred
to a finite Drude peak, while another fraction $\Delta_{\mathrm{AF}}^{2}/\left(\Delta_{\mathrm{AF}}^{2}+\Delta^{2}\right)$
stays at energies above the new optical gap $E_{g}=2\sqrt{\Delta_{\mathrm{AF}}^{2}+\Delta^{2}}$.
Thus, a system without free carriers in the normal state acquires
zero frequency spectral weight through transfer of weight over energies
larger than $2\Delta$. }

\label{fig_transfer} 
\end{figure}

In the parent compounds $\mathrm{BaFe}_{2}\mathrm{As}_{2}$ and $\mathrm{SrFe}_{2}\mathrm{As}_{2}$,
which do not show superconductivity, long range antiferromagnetic
order below the N\'eel temperature $T_{N}$ leads to the formation of
a mid-infrared (MIR) peak\cite{Hu08}, with spectral weight being
transferred to $\omega\simeq2\Delta_{\mathrm{AF}}$ due to the opening
of a spin density wave gap. Here $\Delta_{\mathrm{AF}}$ is the single-particle
gap for momentum states that are Bragg scattered by the magnetic ordering
vector (see figure \ref{fig_band_structure}). For $\mathrm{BaFe}_{2}\mathrm{As}_{2}$,
$2\Delta_{\mathrm{AF}}\simeq1000\mathrm{cm}^{-1}$ (i.e. $\Delta_{\mathrm{AF}}\simeq62\mathrm{meV}$),
yielding $\Delta_{\mathrm{AF}}/\left(k_{B}T_{N}\right)\simeq5$, which
is not unrealistic for itinerant antiferromagnets. For $T<T_{N}$,
the low energy optical response of the parent compounds is characterized
by the Drude form, Eq.(\ref{Drude}), however with a significantly
reduced plasma frequency $\omega_{p,\mathrm{AF}}^{2}\simeq\left(0.1-0.2\right)\omega_{p}^{2}$,
where $\omega_{p}$ is the plasma frequency above $T_{N}$\cite{Hu08}.
Finally, systematic investigations of $\sigma^{\prime}\left(\omega\right)$
in $\mathrm{Ba}(\mathrm{Fe}_{1-x}\mathrm{Co}_{x})_{2}\mathrm{As}_{2}$
as function of temperature and carrier concentration were performed
in Refs.\cite{Uchida10,Lucarelli10}. Upon doping, the MIR peak gets
weaker, consistent with the decrease of the ordered magnetic moment
$M$ with doping\cite{Fernandes10}. Indeed, assuming $\Delta_{\mathrm{AF}}\propto M$
and using the results from Ref.\cite{Fernandes10} for the doping
dependence of $M\left(T=0\right)$, one finds for doping concentrations
where $T_{N}\simeq T_{c}$ that $\Delta_{\mathrm{AF}}\simeq\Delta$,
strongly supporting the view that the same electrons that undergo
Cooper pairing form the ordered moment. This is consistent with the
recent analysis\cite{Fernandes10} of the phase diagram of $\mathrm{Ba}(\mathrm{Fe}_{1-x}\mathrm{Co}_{x})_{2}\mathrm{As}_{2}$:
neutron scattering experiments showed that magnetism and superconductivity
compete strongly, to the extent that the staggered moment is suppressed\cite{Pratt09,Christianson09}
below $T_{c}$ and the magnetic phase boundary $T_{N}\left(x\right)$
is bent back towards smaller $x$-values for $T_{N}<T_{c}$. Theoretical
arguments then demonstrate that this coexistence is only possible
for an $s^{+-}$-pairing state. The optical properties in the regime
of simultaneous magnetic and superconducting order are promising as
they might reveal important information about the interplay between
the superfluid condensate, the normal state Drude peak and the MIR
peak.

In this paper we analyze the optical conductivity in the magnetically
ordered phase as well as in the regime of simultaneous magnetic and
superconducting order. Using parameters suitable to the description
of the phase diagram\cite{Fernandes10} of $\mathrm{Ba}(\mathrm{Fe}_{1-x}\mathrm{Co}_{x})_{2}\mathrm{As}_{2}$,
we numerically obtain the optical spectrum below $T_{N}$. Due to
the partial gapping of the Fermi surface in the itinerant antiferromagnetic
state, we obtain a Drude peak in addition to a MIR peak at $\omega\simeq2\Delta_{\mathrm{AF}}$,
in qualitative agreement with the experimental data. Thus, our results
give further strong evidence for the itinerant character of the
magnetically ordered state.

As shown by nuclear magnetic resonance and muon spin rotation experiments\cite{Laplace09,Julien09,Bernhard09},
antiferromagnetism and superconductivity coexist homogeneously in
$\mathrm{Ba}(\mathrm{Fe}_{1-x}\mathrm{Co}_{x})_{2}\mathrm{As}_{2}$
for the doping range $0.035<x<0.059$. Recent tunnel diode resonator
measurements showed that, in this regime, the $T=0$ superfluid density
is reduced when compared to its value in the pure superconducting
state\cite{Gordon10}. One might expect, at first glance, that such
a reduction is due to the suppressed plasma frequency below $T_{N}$,
and the inverse squared penetration depth $\lambda^{-2}$ is given
by the reduced value $\omega_{p,\mathrm{AF}}^{2}/c^{2}$. In contrast
to this expectation, we find that the superfluid condensate of a superconductor
with magnetic long range order, while reduced compared to the case
without magnetic order, has $\lambda^{-2}$ values that are significantly
larger than $\omega_{p,\mathrm{AF}}^{2}/c^{2}$. In particular, we
find a sizable condensate fraction even in the limit where $\omega_{p,\mathrm{AF}}^{2}=0$.
By analytically investigating the simple but relevant limit of particle-hole
symmetry, we demonstrate that, in the magnetically ordered superconducting
state, spectral weight with energies $\omega\simeq2\Delta_{\mathrm{AF}}$
is transferred from the MIR-peak into the singular $\delta\left(\omega\right)$
term of $\sigma^{\prime}\left(\omega\right)$, enhancing $n_{s}$
(see figure \ref{fig_transfer}). This spectral weight is transferred
from regions of the spectrum that can easily be larger than $2\Delta$,
reflecting the fact that the rigidity of the superconducting ground
state with respect to transverse current fluctuations, while smaller
than in the paramagnet, is still larger than what the low frequency
Drude weight would suggest.

Our results show that the superfluid density in a magnetic superconductor
is not only related to the remaining electronic states of the magnetically
reconstructed Fermi surface at $T=0$, but also to the transfer of
spectral weight around the MIR peak. These conclusions are consistent
with recent theoretical investigations from Vorontsov \emph{et al.}\cite{Vorontsov10}
showing that superconductivity is able to coexist with magnetism even
when the reconstructed Fermi surface at $T=0$ is completely gapped.

The paper is organized as follows: in Section II we review the basic
properties of the optical spectrum of classic superconductors. In
Section III we present our results for the optical conductivity in
the magnetically ordered phase of the iron arsenides. Section IV is
devoted to the investigation of the optical spectrum of the coexistence
state and its relationship to the superfluid density. Section V brings
our conclusions and in Appendix A we present an explicit calculation
of the penetration depth using an alternative approach.

\section{Optical conductivity in superconductors}

Within the Kubo formalism the longitudinal optical conductivity is
given by \begin{equation}
\sigma\left(\omega\right)=\frac{i}{\omega+i0^{+}}\left(\Pi\left(\mathbf{q=0},\omega\right)+\frac{\omega_{p}^{2}}{4\pi}\right)\label{Kubo}\end{equation}
with longitudinal current-current correlation function \begin{equation}
\Pi\left(\mathbf{q},\omega\right)=-i\int_{0}^{\infty}e^{i\omega t}\left\langle \left[j_{\alpha}\left(\mathbf{q,}t\right),j_{\alpha}\left(\mathbf{-q,}0\right)\right]_{-}\right\rangle dt.\end{equation}
where $j_{\alpha}$ is the $\alpha$-th component of the current operator.
For the real part of the optical conductivity, we have: \begin{equation}
\sigma^{\prime}\left(\omega\right)=\ D\delta\left(\omega\right)+\sigma_{\mathrm{reg}}^{\prime}\left(\omega\right),\label{resig}\end{equation}
with the regular contribution $\sigma_{\mathrm{reg}}\left(\omega\right)$,
that does not contain a $\delta\left(\omega\right)$ contribution.
From Eq.(\ref{Kubo}), it follows that the Drude weight $D$ of the
optical conductivity is given by \begin{equation}
D=\frac{1}{4}\left(\omega_{p}^{2}+4\pi\Pi\left(\mathbf{q=0},\omega\rightarrow0\right)\right)\end{equation}
while the regular contribution is \begin{equation}
\sigma_{\mathrm{reg}}^{\prime}\left(\omega\right)=-\frac{\mathrm{Im}\Pi\left(\mathbf{q=0},\omega\right)}{\omega}.\label{sigma_regular}\end{equation}

The current-current correlation function also determines the London
penetration depth via\begin{equation}
\lambda^{-2}=c^{-2}\left(\omega_{p}^{2}+4\pi\Pi\left(\mathbf{q\rightarrow0},\omega=0\right)\right)\label{pen direct}\end{equation}
 i.e. we consider the static current response at small but finite
momentum, in distinction to the weight $D$ that measures the homogeneous
($\mathbf{q=0}$) response at small $\omega$. In general, the order
of the limits $\mathbf{q=0}$, $\omega\rightarrow0$ versus $\omega=0$,
$\mathbf{q\rightarrow0}$, matters and yields different results. Yet,
in the case of a system with gapped excitation spectrum, it was shown
in Ref.\cite{Scalapino93} that the order in which these limits are
taken is irrelevant. Thus, in the case of a fully gapped superconductor,
it follows generally that the Drude weight in the superconductor$\ $
\begin{equation}
D=\frac{c^{2}}{4\lambda^{2}}\label{DD}\end{equation}
 is determined by the penetration depth and thus by the superfluid
condensate \begin{equation}
n_{s}=\frac{m^{\ast}c^{2}}{4\pi e^{2}}\lambda^{-2}.\end{equation}

Formally, a contribution $D\neq0$ in Eq.(\ref{resig}) is not a proof
for superconductivity and may occur in a metallic system that is unable
to relax its momentum. Then, the metal becomes a perfect conductor,
where charges are freely accelerated by an external electric field.
However, in a realistic system one always expects scattering events
that allow for momentum relaxation. In case of a perfect conductor,
such events broaden the singular Drude peak e.g. \begin{equation}
\delta\left(\omega\right)\rightarrow\frac{1}{\pi}\frac{\tau}{1+\left(\omega\tau\right)^{2}}\label{broad}\end{equation}
 with scattering time $\tau$. Thus, formally, $D=0$ and the Drude
response becomes part of the regular contribution to the conductivity
$\sigma_{\mathrm{reg}}^{\prime}\left(\omega\right)$. This is different
for a superconductor, where scattering events may cause a reduction
of the value of $D$ in Eq.(\ref{resig}) but do not change the $\delta\left(\omega\right)$-form
of the zero frequency contribution. The Meissner effect requires$\ $that
$\omega_{p}^{2}>-4\pi\Pi\left(\mathbf{q\rightarrow0},\omega=0\right)$.
Together with the fact that the order of limits does not matter for
a gapped system\cite{Scalapino93}, follows $D>0$, i.e. the $\delta\left(\omega\right)$-form
is robust. This preservation of the singular $\omega=0$ response
is a consequence of the unique rigidity of the superconductor with
respect to transverse current fluctuations. Formally, this rigidity
of the superconducting ground state is reflected by the smallness
of $\left|\Pi\left(\mathbf{q\rightarrow0},\omega=0\right)\right|$
compared to $\omega_{p}^{2}/\left(4\pi\right)$.

A quantitative determination of the penetration depth in a superconductor
can be performed by analyzing the spectral weight transfer in $\sigma\left(\omega\right)$,
and is expressed by the Ferrell-Glover-Tinkham (FGT) sum rule\cite{Ferrell58,Tinkham59}:
\begin{equation}
\lambda^{-2}=\frac{8}{c^{2}}\int_{0^{+}}^{\infty}\left(\sigma_{ns}^{\prime}\left(\omega\right)-\sigma_{sc}^{\prime}\left(\omega\right)\right)d\omega.\label{FGT}\end{equation}

This sum rule relates $\lambda$ to the change of total spectral weight
between the normal state ($ns$) and the superconducting ($sc$) state
for $\omega>0$. It follows from the $f$-sum rule, Eq. (\ref{fsum}),
and the emergence of the $D\delta\left(\omega\right)$-term only below
$T_{c}$. Thus, upon entering the superconducting state, spectral
weight is transferred from finite frequencies to the $\delta$-function
at $\omega=0$. We mention that for Eq.(\ref{FGT}) to hold, one assumes
that the expectation value of the optical mass $m^{\ast}$, i.e. the
value of $\omega_{p}^{2}=4\pi e^{2}n/m^{\ast}$, is unaffected by
the onset of superconductivity. In the FeAs superconductors, this
seems to be the case\cite{Li08,Uchida10}, in distinction to the evidence
for violation of the FGT sum rule in cuprate superconductors\cite{Basov05}.

\section{Optical spectrum in the itinerant antiferromagnetic phase}

\subsection{Microscopic model for competing magnetic and superconducting order}

To develop a microscopic model of the interplay between superconductivity
and magnetism, we use a few basic ingredients to describe the main
features of the iron arsenides\cite{Fernandes10}: the electronic
structure is characterized by two sets of Fermi surface sheets, a
circular hole pocket around the center of the Brillouin zone and an
elliptical electron pocket shifted by the magnetic ordering vector
$\mathbf{Q}$. The non-interacting part $\mathcal{H}_{0}$ of the
Hamiltonian is then given by

\begin{figure}

\begin{centering}
\includegraphics[width=1\columnwidth]{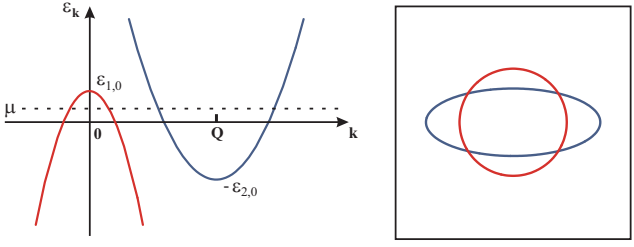} 
\par\end{centering}

\caption{Schematic representation of the band structure considered here (left
panel).The Fermi surface has an elliptical electron pocket (blue ellipse)
displaced by the magnetic ordering vector $\mathbf{Q}$ from the circular
hole pocket (red circle). In the right panel, we present the Fermi
surface at zero doping. For simplicity, we displaced the centers of
the bands to make them coincide. }

\label{fig_band_structure} 
\end{figure}

\begin{equation}
\mathcal{H}_{0}=\sum_{\mathbf{k}\sigma}\left(\xi_{1,\mathbf{k}}c_{\mathbf{k}\sigma}^{\dagger}c_{\mathbf{k}\sigma}+\xi_{2,\mathbf{k}}d_{\mathbf{k}\sigma}^{\dagger}d_{\mathbf{k}\sigma}\right).\label{H0}\end{equation}

We consider only one hole band located in the center of the Brillouin
zone with dispersion $\xi_{1,\mathbf{k}}$, and one electron band,
shifted by $\mathbf{Q}$ from the hole band, with dispersion $\xi_{2,\mathbf{k}}$
(see figure \ref{fig_band_structure}):

\begin{eqnarray}
\xi_{1,\mathbf{k}} & = & \varepsilon_{1,0}-\frac{k^{2}}{2m}-\mu\\
\xi_{2,\mathbf{k+Q}} & = & -\varepsilon_{2,0}+\frac{k_{x}^{2}}{2m_{x}}+\frac{k_{y}^{2}}{2m_{y}}-\mu,\label{band_dispersion}\end{eqnarray}

A magnetic interaction $I$ and an interband pairing interaction $V$
lead to the possibility of antiferromagnetic order with antiferromagnetic
gap\begin{equation}
\Delta_{\mathrm{AF}}=\frac{I}{2}\sum_{\mathbf{k},\sigma}\sigma\left\langle c_{\mathbf{k}\sigma}^{\dagger}d_{\mathbf{k+Q}\sigma}\right\rangle \end{equation}
 and superconductivity with coupled gap equations \begin{eqnarray}
\Delta_{2} & = & -\frac{V}{N}\sum_{\mathbf{k}}\left\langle c_{\mathbf{k}\uparrow}^{\dagger}c_{-\mathbf{k}\downarrow}^{\dagger}\right\rangle \\
\Delta_{1} & = & -\frac{V}{N}\sum_{\mathbf{k}}\left\langle d_{\mathbf{k+Q}\uparrow}^{\dagger}d_{-\mathbf{k}-\mathbf{Q}\downarrow}^{\dagger}\right\rangle .\end{eqnarray}

For $V<0$, as it would be the case for phonon mediated superconductivity,
one obtains the $s^{++}$ state ($\Delta_{1}\Delta_{2}>0$), whereas
for $V>0$ it follows the unconventional sign-changing $s^{+-}$ state
($\Delta_{1}\Delta_{2}<0$). Introducing the Nambu operator $\Psi_{\mathbf{k}}=\left(c_{\mathbf{k\uparrow}},c_{-\mathbf{k\downarrow}}^{\dagger},d_{\mathbf{k+Q\uparrow}},d_{-\mathbf{k}-\mathbf{Q}\mathbf{\downarrow}}^{\dagger}\right)^{T}$,
we consider the mean field Hamiltonian\begin{equation}
H=\int_{\mathbf{k}}\Psi_{\mathbf{k}}^{\dagger}\widehat{\varepsilon}_{\mathbf{k}}\Psi_{\mathbf{k}}\end{equation}
 with simultaneous antiferromagnetic and superconducting order. Here
\begin{equation}
\widehat{\varepsilon}_{\mathbf{k}}=\left(\begin{array}{cccc}
\xi_{1,\mathbf{k}} & \Delta_{1} & \Delta_{\mathrm{AF}} & 0\\
\Delta_{1} & -\xi_{1,\mathbf{k}} & 0 & \Delta_{\mathrm{AF}}\\
\Delta_{\mathrm{AF}} & 0 & \xi_{2,\mathbf{k+Q}} & \Delta_{2}\\
0 & \Delta_{\mathrm{AF}} & \Delta_{2} & -\xi_{2,\mathbf{k+Q}}\end{array}\right)\end{equation}

For details of this model, see Refs \cite{Fernandes10,Fernandes10_2}.
Now, the $\mathbf{q=0}$ current-current correlation function is given
by \begin{equation}
\Pi_{\alpha\beta}\left(i\omega_{n}\right)=e^{2}T\sum_{\mathbf{k},m}\mathrm{tr}\left(\widehat{v}_{\mathbf{k}\alpha}\widehat{G}_{\mathbf{k}}\left(i\nu_{m}+i\omega_{n}\right)\widehat{v}_{\mathbf{k}\beta}\widehat{G}_{\mathbf{k}}\left(i\nu_{m}\right)\right)\label{current_current}\end{equation}
 with $\widehat{G}_{\mathbf{k}}^{-1}\left(i\omega_{n}\right)=i\omega_{n}\widehat{1}-\widehat{\varepsilon}_{\mathbf{k}}$
and the velocity matrix in Nambu space $\widehat{v}_{\mathbf{k}\alpha}=\partial/\partial k_{\alpha}\mathrm{diag}\left(\xi_{1,\mathbf{k}},\xi_{1,\mathbf{k}},\xi_{2,\mathbf{k+Q}},\xi_{2,\mathbf{k+Q}}\right)$.
Here, the indices $\alpha$ and $\beta$ refer to Cartesian components of vectors, $\omega_n=2n \pi T$ is a bosonic Matsubara frequency and $\nu_m = (2m+1)\pi T$ is a fermionic Matsubara frequency.

\subsection{Magnetically ordered phase without superconductivity}

First, we investigate the optical properties of the pure antiferromagnetic
state. In this case, one has $2\times2$ matrices in Nambu space and
the Green's function is given by:

\begin{eqnarray}
\widehat{G}_{s\mathbf{k}}\left(i\omega_{n}\right) & = & \left(i\omega_{n}-E_{1,\mathbf{k}}\right)^{-1}\left(i\omega_{n}-E_{2,\mathbf{k}}\right)^{-1}\times\nonumber \\
 &  & \left(\begin{array}{cc}
i\omega_{n}-\xi_{2,\mathbf{k+Q}} & -s\Delta_{\mathrm{AF}}\\
-s\Delta_{\mathrm{AF}} & i\omega_{n}-\xi_{1,\mathbf{k}}\end{array}\right)\label{green_function_AFM}\end{eqnarray}
where $s$ denotes the spin and $E_{a,\mathbf{k}}$, the quasiparticle
energy:

\begin{equation}
E_{a,\mathbf{k}}=\left(\frac{\xi_{1,\mathbf{k}}+\xi_{2,\mathbf{k+Q}}}{2}\right)\pm\sqrt{\Delta_{\mathrm{AF}}^{2}+\left(\frac{\xi_{1,\mathbf{k}}-\xi_{2,\mathbf{k+Q}}}{2}\right)^{2}}\label{quasiparticle_AFM}\end{equation}

To evaluate the current-current correlation function, Eq. (\ref{current_current}),
we use the Kramers-Kronig relations:

\begin{equation}
\widehat{G}_{s\mathbf{k}}\left(i\omega_{n}\right)=-\int_{-\infty}^{\infty}\frac{d\Omega}{\pi}\frac{\mathrm{Im}\widehat{G}_{s\mathbf{k}}\left(\Omega+i0^{+}\right)}{i\omega_{n}-\Omega}\label{Kramers_Kronig}\end{equation}

In order to obtain more realistic results, we replace one of the $i0^{+}$
convergence factors above by a finite single-particle lifetime $i\tau^{-1}$.
Then, the real part of the conductivity is given only by the regular
contribution (\ref{sigma_regular}). A straightforward calculation
leads to:

\begin{eqnarray}
\sigma_{\alpha\alpha}^{\prime}\left(\omega\right) & = & 2e^{2}v_{F,\alpha}^{2}\left(\frac{\tau^{-1}}{\omega^{2}+\tau^{-2}}\right)\sum_{a=1}^{2}\sum_{\mathbf{k}}\frac{f_{\alpha}\left(E_{a,\mathbf{k}},\omega\right)}{\left(E_{a,\mathbf{k}}-E_{\bar{a},\mathbf{k}}\right)}\times\nonumber \\
 &  & \frac{\left(2\omega+E_{a,\mathbf{k}}-E_{\bar{a},\mathbf{k}}\right)}{\left(\omega+E_{a,\mathbf{k}}-E_{\bar{a},\mathbf{k}}\right)^{2}+\tau^{-2}}\end{eqnarray}
 where $\mathbf{v}_{F}$ is the hole-band Fermi velocity and:

\begin{eqnarray}
f_{\alpha}\left(E_{a,\mathbf{k}},\omega\right) & = & \left[\frac{n_{F}\left(E_{a,\mathbf{k}}\right)-n_{F}\left(E_{a,\mathbf{k}}+\omega\right)}{\omega}\right]\times\nonumber \\
 &  & \left(-\frac{2\Delta_{\mathrm{AF}}^{2}}{\bar{m}_{\alpha}}+\frac{C_{1,\mathbf{k}}^{(a)}}{\bar{m}_{\alpha}^{2}}+C_{2,\mathbf{k}}^{(a)}\right)\end{eqnarray}
 with $C_{i,\mathbf{k}}^{(a)}=\left(E_{a,\mathbf{k}}-\xi_{i,\mathbf{k}}\right)\left(E_{a,\mathbf{k}}+\omega-\xi_{i,\mathbf{k}}\right)$,
Fermi function $n_{F}$ and relative electron-band mass $\bar{m}_{\alpha}=m_{\alpha}/m$. Here, we introduced the index $\bar{a}$ defined as $\bar{a}=2$ for $a=1$ and $\bar{a}=1$ for $a=2$. 

\begin{figure}

\begin{centering}
\includegraphics[width=0.8\columnwidth]{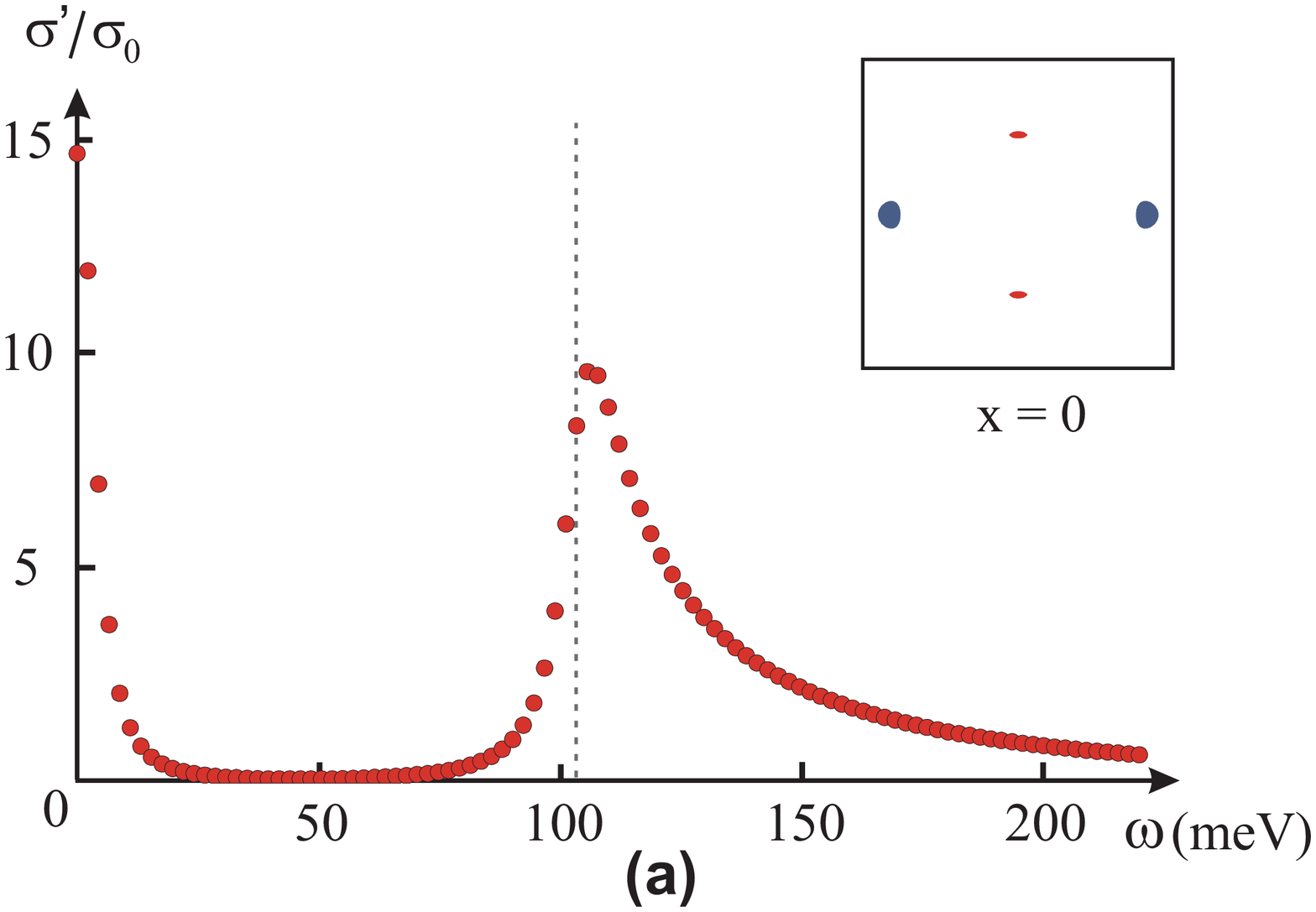} 
\par\end{centering}

\begin{centering}
\includegraphics[width=0.8\columnwidth]{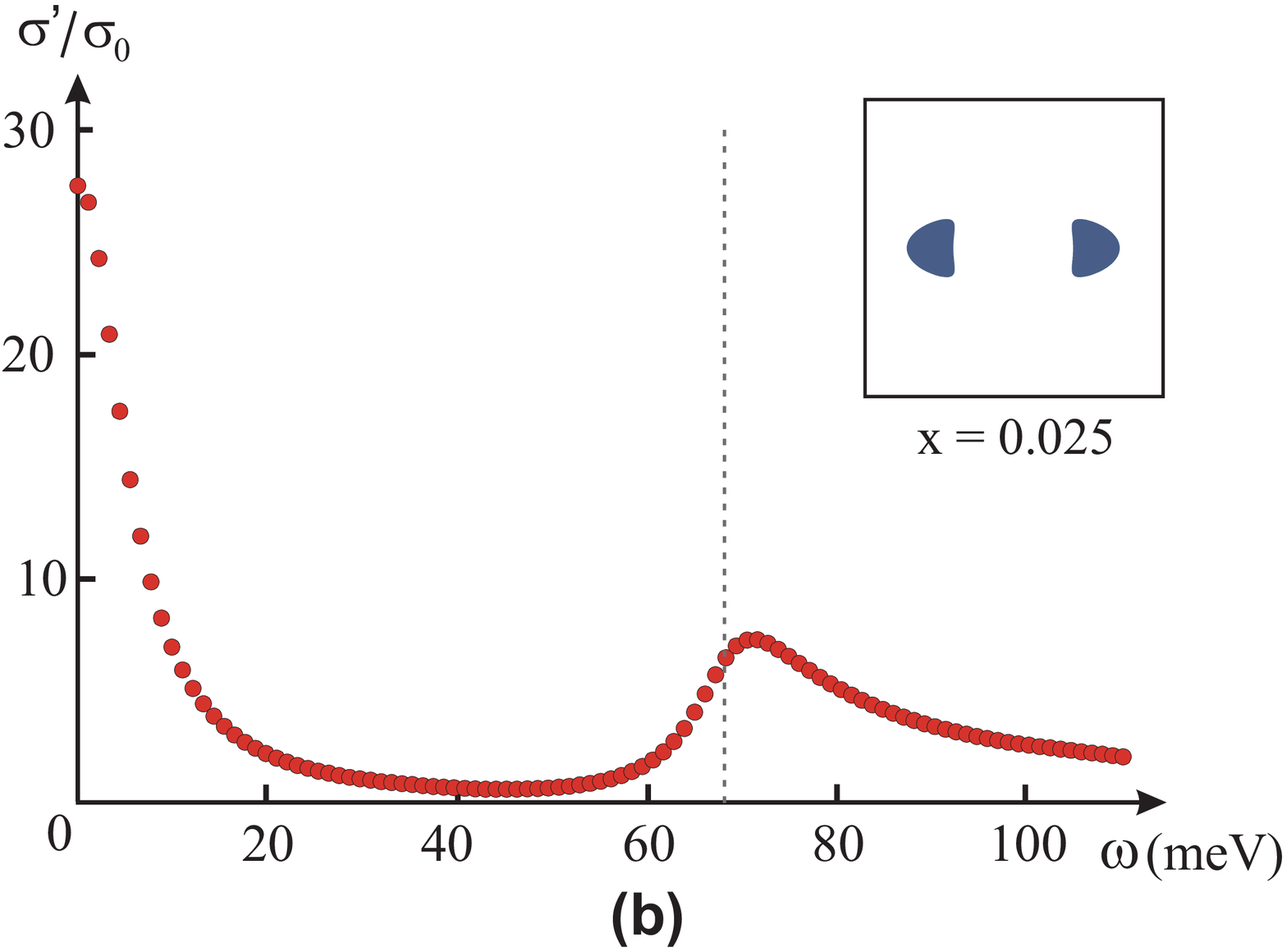} 
\par\end{centering}

\begin{centering}
\includegraphics[width=0.8\columnwidth]{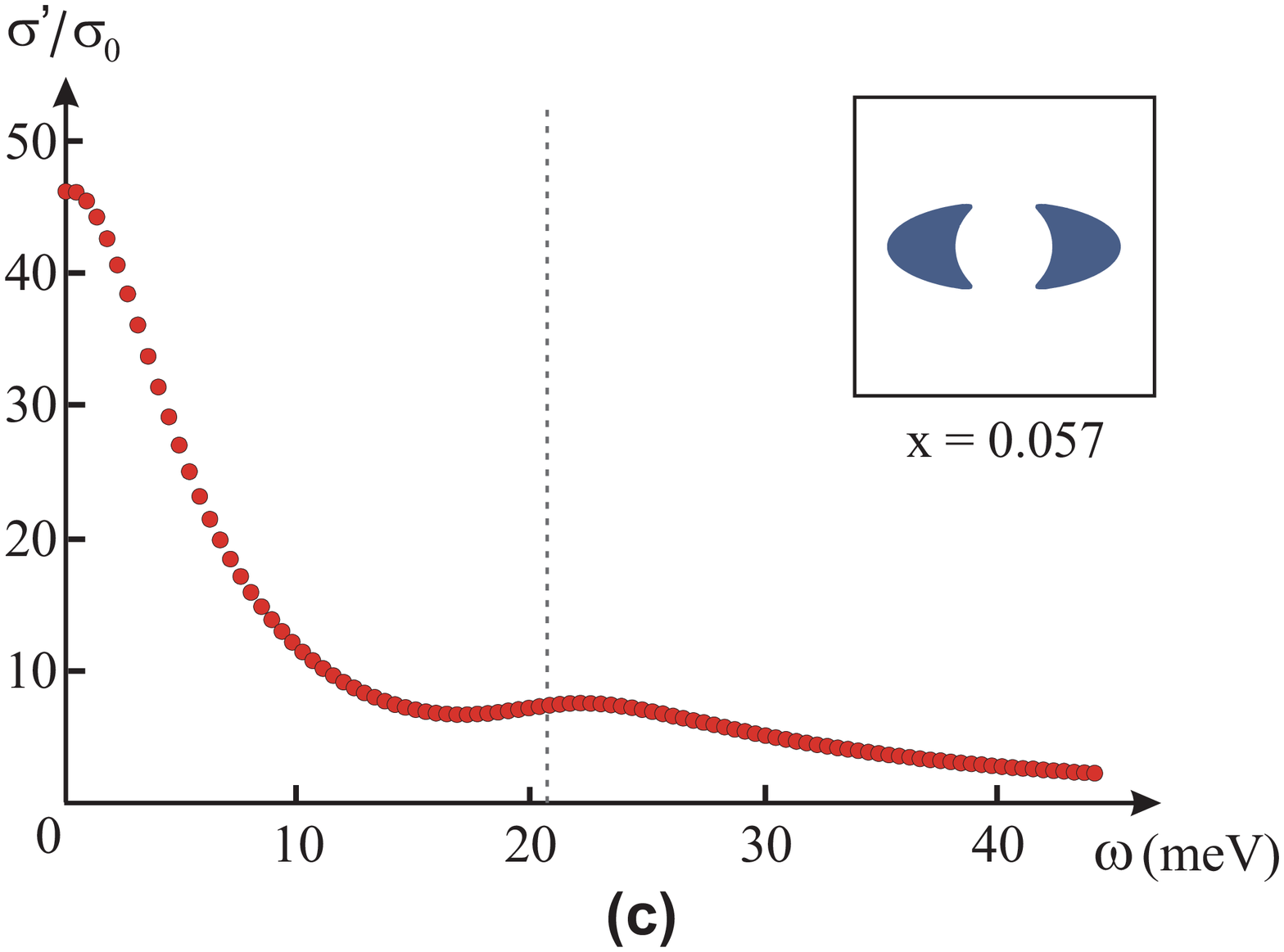} 
\par\end{centering}

\caption{Real part of the optical conductivity $\sigma^{\prime}$ (in units
of $\sigma_{0}=2\pi^{-1}\rho e^{2}v_{F,x}^{2}/\varepsilon_{0}$) as
function of the frequency $\omega$ (in units of meV) at $T\approx0$
for different Co doping concentrations: (a) $x=0$, (b) $x=0.025$,
and (c) $x=0.057$. We use the band structure parameters that consistently
describe the phase diagram of $\mathrm{Ba}(\mathrm{Fe}_{1-x}\mathrm{Co}_{x})_{2}\mathrm{As}_{2}$
(see main text for more details); only magnetic ordering is considered.
The dashed gray line refers to the frequency $2\Delta_{\mathrm{AF}}$
and the inset shows the magnetically reconstructed Fermi surface (see
figure \protect\ref{fig_band_structure}). }

\label{fig_conduct_AFM} 
\end{figure}

In Ref. \cite{Fernandes10}, we introduced the band structure parameters
that provide a good agreement between the model of the previous subsection
and the neutron diffraction data on $\mathrm{Ba}(\mathrm{Fe}_{1-x}\mathrm{Co}_{x})_{2}\mathrm{As}_{2}$.
Specifically, they are given by $\varepsilon_{1,0}=0.095$ eV, $\varepsilon_{2,0}=0.125$ eV,
$m=1.32m_{\mathrm{electron}}$, $m_{x}=2m$, and $m_{y}=0.3m$, together
with the electronic interaction $I=0.95$ eV and the assumption that
each Co atom adds one extra electron. In this subsection, since we
are only interested in the pure antiferromagnetic phase, we set $V=0$.
In figure \ref{fig_conduct_AFM}, we use these parameters to obtain
the real part of the optical conductivity at approximately zero temperature
and for different Co doping concentrations, with no superconductivity
involved. Note that the only free parameter here is the single-particle
lifetime $\tau$, which was chosen to be $\tau^{-1}=5.5$ meV for
all doping concentrations. This is the same order of magnitude of
the scattering rate associated with the narrow Drude peak observed in
optical experiments\cite{Lucarelli10}.

Our objective here is not to describe all the details of the observed
optical spectrum, which would require a description of all five Fe-3d
orbitals (see, for example, Ref. \cite{Kaneshita09}), but rather to understand its main features. From figure
\ref{fig_conduct_AFM}, we see that, in general, the optical spectrum
in the antiferromagnetic phase has a Drude peak as well as a finite-frequency
peak. The latter is located very close to $\omega\simeq2\Delta_{\mathrm{AF}}$,
and is associated with the opening of the spin density wave gap. As
shown in the same figure, $\Delta_{\mathrm{AF}}$ only partially gaps
the Fermi surface, resulting in a finite Drude peak proportional to
$\tau$. Recall that, for nested bands and $\tau\rightarrow\infty$,
the reconstructed Fermi surface is completely gapped and the optical
conductivity in the magnetically ordered phase vanishes for $\omega<2\Delta_{\mathrm{AF}}$
(see Eq. \ref{sig reg res} below for $\Delta=0$).

The existence of a peak at $2\Delta_{\mathrm{AF}}$, combined with
the $f$-sum rule, Eq. (\ref{fsum}), implies that the plasma frequency
associated with the remaining Drude peak in the antiferromagnetic state
must be smaller than the plasma frequency of the Drude peak in the
paramagnetic phase, as seen experimentally\cite{Uchida10} (assuming
that the optical mass is the same in both situations). Note that the
theoretical value of $\Delta_{\mathrm{AF}}$ in the undoped sample
is $\Delta_{\mathrm{AF}}\approx51$ meV (Fig. \ref{fig_conduct_AFM}a),
which is very close to the value extracted from the measured optical
spectrum\cite{Hu08}. As doping increases and the magnitude of the
gap decreases, the MIR peak gets weaker and moves towards lower frequencies
(Fig. \ref{fig_conduct_AFM}b), until it is almost completely masked
by the Drude peak (Fig. \ref{fig_conduct_AFM}c). These results are
in general agreement with optical measurements\cite{Hu08,Uchida10,Lucarelli10}
on $\mathrm{Ba}(\mathrm{Fe}_{1-x}\mathrm{Co}_{x})_{2}\mathrm{As}_{2}$,
demonstrating the itinerant character of the magnetically ordered
state in these compounds.

\section{Optical spectrum in the magnetically ordered superconducting phase}

So far we have considered only the magnetically ordered state. However,
for $\mathrm{Ba}(\mathrm{Fe}_{1-x}\mathrm{Co}_{x})_{2}\mathrm{As}_{2}$,
superconductivity coexists with antiferromagnetism at very low temperatures
for $0.035<x<0.059$. In order to achieve a more transparent insight
about the optical conductivity in the coexistence state, we investigate
analytically the limit of particle-hole symmetry, where $\varepsilon_{0}\equiv\varepsilon_{1,0}=\varepsilon_{2,0}$,
$m_{x}=m_{y}=m$ and $\mu=0$. In this case, $\xi_{\mathbf{k}}\equiv\xi_{1,\mathbf{k}}=-\xi_{2,\mathbf{k+Q}}$,
implying that the hole and electron Fermi surfaces are identical (perfect
nesting). In the case of $s^{+-}$-pairing we have \begin{equation}
\widehat{\varepsilon}_{\mathbf{k}}=\xi_{\mathbf{k}}\tau_{3}\sigma_{3}+\Delta_{\mathrm{AF}}\tau_{0}\sigma_{1}+\Delta\tau_{1}\sigma_{z}\label{energy}\end{equation}
 where $\Delta=\Delta_{1}=-\Delta_{2}$ and $\tau_{\alpha}$ and $\sigma_{\beta}$
are the Pauli matrices that act in Nambu and band space, respectively.
In case of $s^{++}$-pairing, we replace $\sigma_{z}$ by $\sigma_{0}$
in the last term. Eq.(\ref{energy}) leads to the single particle
Green's function \begin{equation}
\widehat{G}_{\mathbf{k}}\left(i\omega_{n}\right)=-\frac{i\omega_{n}\tau_{0}\sigma_{0}+\widehat{\varepsilon}_{\mathbf{k}}}{\omega_{n}^{2}+\Delta^{2}+\Delta_{\mathrm{AF}}^{2}+\xi_{\mathbf{k}}^{2}}.\label{greens_function_phs}\end{equation}

At zero temperature the $\mathbf{q=0}$ current-current 
correlation function is: \begin{equation}
\Pi\left(i\omega\right)=\frac{\omega_{p}^{2}}{2}\int\frac{d\Omega d\xi}{\left(4\pi\right)^{2}}\:\mathrm{tr}\left(\widehat{v}_{0}\widehat{G}_{\mathbf{k}}\left(i\omega+i\Omega\right)\widehat{v}_{0}\widehat{G}_{\mathbf{k}}\left(i\Omega\right)\right)\end{equation}
 where $\widehat{v}_{0}=\tau_{0}\sigma_{z}$ contains the proper sign
of the current vertex. Performing the trace over the band and Nambu
degrees of freedom, we find \begin{eqnarray}
\Pi\left(\omega\right) & = & \frac{\omega_{p}^{2}}{8\pi^{2}}\int d\Omega d\xi\frac{\xi^{2}+\Delta^{2}-\Delta_{\mathrm{AF}}^{2}-\Omega\left(\omega+\Omega\right)}{\xi^{2}+\Delta_{\mathrm{AF}}^{2}+\Delta^{2}+\Omega^{2}}\nonumber \\
 &  & \times\frac{1}{\xi^{2}+\Delta_{\mathrm{AF}}^{2}+\Delta^{2}+\left(\omega+\Omega\right)^{2}}\label{Pii}\end{eqnarray}

In the limit $\omega\rightarrow0$ follows \begin{equation}
\Pi\left(\omega\rightarrow0\right)=-\frac{\omega_{p}^{2}}{4\pi}\frac{\Delta_{\mathrm{AF}}^{2}}{\Delta_{\mathrm{AF}}^{2}+\Delta^{2}}\end{equation}
 and we obtain for the optical conductivity \begin{equation}
D=\frac{\omega_{p}^{2}}{4}\frac{\Delta^{2}}{\Delta_{\mathrm{AF}}^{2}+\Delta^{2}}.\end{equation}

For the non-superconducting antiferromagnet ($\Delta=0$ but $\Delta_{\mathrm{AF}}\neq0$)
follows $D=0$. This is a consequence of perfect nesting that leads
to a fully gapped antiferromagnetic state. In the non-magnetic superconductor
($\Delta_{\mathrm{AF}}=0$ but $\Delta\neq0$) the current-current
correlation function vanishes at $T=0$. Then, it follows $D=\omega_{p}^{2}/4$
and Eq.(\ref{DD}) yields the BCS result for the penetration depth
$\lambda_{0}=\omega_{p}^{2}/c^{2}$. In the general case we obtain
for the penetration depth \begin{equation}
\lambda^{-2}=\lambda_{0}^{-2}\:\frac{\Delta^{2}}{\Delta_{\mathrm{AF}}^{2}+\Delta^{2}}\label{pen res}\end{equation}

We point out that this result is the same as in the case of a charge
density wave state coexisting with a conventional $s$-wave state\cite{Machida84}.
As shown in Appendix A, we obtain the same result for the penetration
depth by explicitly analyzing Eq.(\ref{pen direct}), i.e. by first
taking $\omega=0$ and then $\mathbf{q}\rightarrow0$. In this context,
the existence of a finite $\Pi\left(\omega=0,\mathbf{q}\rightarrow0\right)$
at $T=0$ is related to the fact that one of the two coherence factors
is not identically zero, in contrast to what happens for non-magnetic
superconductors. Thus, the rigidity of the non-magnetic BCS ground
state with respect to transverse current fluctuations is reduced in
the magnetically ordered state.

Note that, formally, the coexistence between superconductivity and
magnetism is only marginal for particle-hole symmetry, as we discussed
elsewhere\cite{Fernandes10}. Yet, small perturbations in both the
chemical potential and the ellipticity of the electron band are able
to place the system in the coexistence regime\cite{Fernandes10,Vavilov09,Vorontsov10,Fernandes10_2}.
Following Refs. \cite{Vavilov09,Vorontsov10}, we can investigate
the effect of these small perturbations on our result for the penetration
depth (\ref{pen res}) by considering the perturbed band structure
$\xi_{2,\mathbf{k}\mathbf{+Q}}=-\xi_{1,\mathbf{k}}-2\delta_{\varphi}$,
with $\delta_{\varphi}=\delta_{0}+\delta_{2}\cos2\varphi$ such that
$\delta_{0}=\mu+\frac{k_{F}^{2}}{4}\left(\frac{1}{m}-\frac{m_{x}+m_{y}}{2m_{x}m_{y}}\right)$
and $\delta_{2}=\frac{k_{F}^{2}}{8}\left(\frac{m_{x}-m_{y}}{m_{x}m_{y}}\right)$.
Here, $\varphi$ is the angle on the elliptical electron pocket. A
straightforward calculation leads to:

\begin{equation}
\lambda^{-2}\approx\lambda_{0}^{-2}\:\frac{\Delta^{2}}{\Delta_{\mathrm{AF}}^{2}+\Delta^{2}}\left[1+\frac{4\Delta_{\mathrm{AF}}^{2}}{3\left(\Delta_{\mathrm{AF}}^{2}+\Delta^{2}\right)^{2}}\left\langle \delta_{\varphi}^{2}\right\rangle \right]\label{away_phsymmetry}\end{equation}

\begin{figure}

\begin{centering}
\includegraphics[width=0.85\columnwidth]{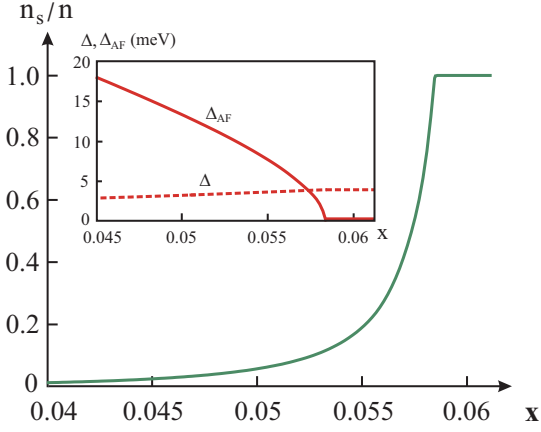} 
\par\end{centering}

\caption{Ratio of the superfluid condensate $n_{s}$ and the electron density
$n$ with parameters relevant for $\mathrm{Ba}(\mathrm{Fe}_{1-x}\mathrm{Co}_{x})_{2}\mathrm{As}_{2}$
in the clean limit, as function of $x$ and for $T=0$. For $x\lesssim0.059$
simultaneous antiferromagnetic and superconducting order sets in while
for $x\gtrsim0.059$ no long range magnetic order exists. The inset shows the $T=0$ values of $\Delta$ (dashed red line)
and $\Delta_{\mathrm{AF}}$ (solid red line) as function of $x$, according to the theory
of Ref. \cite{Fernandes10}. For simplicity we consider the particle-hole
symmetry expression (\ref{pen res}) and use the value of the SC gap referent to the electron band.}

\label{fig_penetration_depth} 
\end{figure}
with $\left\langle \delta_{\varphi}^{2}\right\rangle =\delta_{0}^{2}+\delta_{2}^{2}/2$.
Thus, both perturbations in the chemical potential and in the ellipticity
lead to a decrease in the penetration depth, i.e. to an increase in
the value of $D$. Therefore, we can interpret the particle-hole symmetric
result (\ref{pen res}) as an {}``upper-bound'' for $\lambda$.
With this in mind, even though the band structure of the pnictides
is not particle-hole symmetric, it is instructive to substitute in
Eq. (\ref{pen res}) the $T=0$ values of $\Delta$ and $\Delta_{\mathrm{AF}}$
obtained by numerically solving the gap equations with the band structure
parameters of $\mathrm{Ba}(\mathrm{Fe}_{1-x}\mathrm{Co}_{x})_{2}\mathrm{As}_{2}$
(see Section III). The results are displayed in figure \ref{fig_penetration_depth}.
Remarkably, similar values for the relative increase of $\lambda_{T=0}$
have been recently measured by Gordon \emph{et al}.\cite{Gordon10} along
the coexistence region of $\mathrm{Ba}(\mathrm{Fe}_{1-x}\mathrm{Co}_{x})_{2}\mathrm{As}_{2}$
using the tunnel diode resonator technique.

Next, we analyze the optical conductivity at finite frequencies. To
this end, we evaluate the integrations in Eq.(\ref{Pii}) and perform
the analytical continuation to the real frequency axis, $i\omega\rightarrow\omega+i0^{+}$,
obtaining: \begin{equation}
\sigma_{reg}^{\prime}\left(\omega\right)=\left\{ \begin{array}{cc}
0 & \omega<E_{g}\\
\frac{\omega_{p}^{2}}{2\omega^{2}}\frac{\Delta_{\mathrm{AF}}^{2}}{\sqrt{\omega^{2}-E_{g}^{2}}} & \omega\geq E_{g}\end{array}\right.\label{sig reg res}\end{equation}
 with the optical gap \begin{equation}
E_{g}=2\sqrt{\Delta_{\mathrm{AF}}^{2}+\Delta^{2}}\end{equation}

In the normal state, $\Delta=0$ and the optical conductivity is nonzero
only for $\omega>$ $2\Delta_{\mathrm{AF}}$. Entering the superconducting
state, it follows for $\Delta<\Delta_{\mathrm{AF}}$ that there is
no spectral weight in the normal state for $\omega<2\Delta$. Thus,
the finite penetration depth obtained in Eq.(\ref{pen res}) must
be due to the transfer of spectral weight that involves energies above
$2\Delta$. Indeed, analyzing the remaining high frequency spectral
weight, we find from Eq.(\ref{sig reg res}) that

\begin{equation}
2\int_{E_{g}}^{\infty}\sigma_{reg}^{\prime}\left(\omega\right)d\omega=\frac{\omega_{p}^{2}}{4}\frac{\Delta_{\mathrm{AF}}^{2}}{\Delta_{\mathrm{AF}}^{2}+\Delta^{2}}\end{equation}
where the factor of $2$ accounts for negative frequencies. Thus,
the total weight of the non-superconducting antiferromagnet splits
in two parts with ratio $\left(\Delta/\Delta_{\mathrm{AF}}\right)^{2}$.
Below $T_{c}$ a fraction $\ \Delta^{2}/\left(\Delta_{\mathrm{AF}}^{2}+\Delta^{2}\right)$
is transferred to $\omega=0$ to yield Meissner screening and a finite
penetration depth. In addition, the fraction $\Delta_{\mathrm{AF}}^{2}/\left(\Delta_{\mathrm{AF}}^{2}+\Delta^{2}\right)$
remains at energies above the optical gap $E_{g}$, as illustrated
in figure \ref{fig_transfer}. Here we use the fact that $E_{g}$
of the non-superconducting antiferromagnetic state and of the state
with $\Delta$ and $\Delta_{\mathrm{AF}}$ finite is essentially the
same due to the reduced ordered moment below $T_{c}$, see Ref.\cite{Fernandes10}.

Analogously, one can rationalize this result using Ferrell-Glover-Tinkham
(FGT) sum rule, Eq. (\ref{FGT}). By calculating the difference of
spectral weight between the non-superconducting state and the superconducting
state with the same value of $\Delta_{\mathrm{AF}}$, we obtain exactly
the Drude weight $D$:

\begin{eqnarray}
 &  & 2\int_{0^{+}}^{\infty}\left[\sigma_{reg}^{\prime}\left(\omega,\Delta_{\mathrm{AF}},\Delta=0\right)-\sigma_{reg}^{\prime}\left(\omega,\Delta_{\mathrm{AF}},\Delta\right)\right]d\omega\nonumber \\
 &  & =\frac{\omega_{p}^{2}}{4}\frac{\Delta^{2}}{\Delta_{\mathrm{AF}}^{2}+\Delta^{2}}\label{FGT_mag}\end{eqnarray}

Our analysis for particle-hole symmetry clearly shows that, in a magnetic
superconductor, the transfer of spectral weight to the Drude peak
below $T_{c}<T_{N}$ is not from $\omega<2\Delta$, but from higher
frequencies $\omega<2\sqrt{\Delta_{\mathrm{AF}}^{2}+\Delta^{2}}$.
Therefore, analyzing our results for the optical conductivity of the
pure magnetic phase (figure \ref{fig_conduct_AFM}), as well as the
experimental optical spectrum, the superfluid condensate is formed
not only by the reduced remaining Drude peak but also by the significant
portion of spectral weight associated with the MIR peak. More importantly,
even in the absence of a remaining Drude peak in the pure magnetic
state at $T=0$, spectral weight can be transferred to a $\delta\left(\omega\right)$-term
below $T_{c}$, yielding a finite superfluid density.

\begin{figure}

\begin{centering}
\includegraphics[width=0.85\columnwidth]{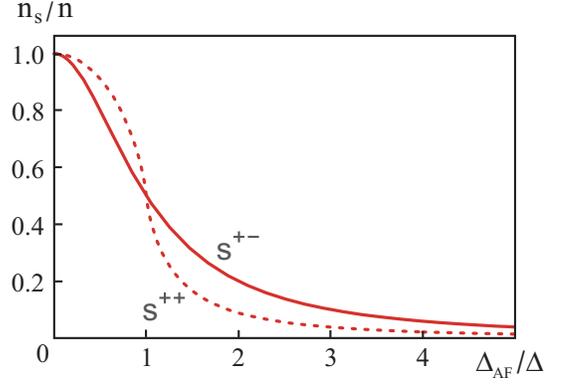} 
\par\end{centering}

\caption{Ratio of the superfluid density $n_{s}$ and the electron density
$n$ as function of the antiferromagnetic gap $\Delta_{\mathrm{AF}}$
(in units of the superconducting gap $\Delta$) for $s^{++}$-pairing
(dashed line) and $s^{+-}$-pairing (solid line) at $T=0$.}

\label{fig_s_plusplus} 
\end{figure}

Let us briefly discuss the situation for $s^{++}$-superconductivity
and particle-hole symmetric bands. Similar calculations lead to the $T=0$
penetration depth \begin{equation}
\lambda^{-2}=\lambda_{0}^{-2}\frac{2\Delta\Delta_{\mathrm{AF}}+\left(\Delta_{\mathrm{AF}}^{2}-\Delta^{2}\right)\ln\left\vert \frac{\Delta_{\mathrm{AF}}-\Delta}{\Delta_{\mathrm{AF}}+\Delta}\right\vert }{4\Delta\Delta_{\mathrm{AF}}}.\end{equation}

Correspondingly, the high frequency contribution has the total weight
$1-\left(\lambda_{0}/\lambda\right)^{2}$ . In figure \ref{fig_s_plusplus}
we plot the superfluid density $n_{s}\left(\Delta,\Delta_{\mathrm{AF}}\right)$ as
function of $\Delta_{\mathrm{AF}}/\Delta$ for $s^{++}$ and $s^{+-}$-pairing.
We see that for $\Delta_{\mathrm{AF}}<\Delta$ holds that $n_{s}$
is larger for $s^{++}$-pairing compared to $s^{+-}$, while the opposite
is true for $\Delta_{\mathrm{AF}}>\Delta$. Qualitatively, the two
behaviors are not very different and would not provide a sharp criterion
to identify the symmetry of the pairing state in the iron arsenides.
Yet, previous analysis\cite{Fernandes10,Fernandes10_2,Vorontsov10}
demonstrated that the conventional $s^{++}$ state is unable to coexist
with itinerant magnetism in the pnictides.

\section{Conclusions}

In conclusion, we analyzed the optical conductivity of both an itinerant
antiferromagnetic state and a magnetically ordered superconductor.
For the pure magnetic phase, using the parameters associated to the
phase diagram\cite{Fernandes10,Fernandes10_2} of $\mathrm{Ba}(\mathrm{Fe}_{1-x}\mathrm{Co}_{x})_{2}\mathrm{As}_{2}$,
we were able to identify the main features observed in the experimental
optical spectrum\cite{Hu08,Uchida10,Lucarelli10}. In particular,
for the undoped compound, we found a reduced Drude peak associated
to the remaining reconstructed parts of the Fermi surface, as well
as a mid-infrared (MIR) peak at $\omega\simeq2\Delta_{\mathrm{AF}}\approx51$
meV, associated to the gap opened at momentum $\mathbf{k_{0}}$ that
are Bragg scattered by the magnetic ordering vector $\mathbf{Q}$,
i.e. $\xi_{1,\mathbf{k}_{0}}=\xi_{2,\mathbf{k_{0}+Q}}$. Upon doping,
the spin density wave gap is reduced and, consequently, the MIR peak
becomes weaker and more masked by the Drude peak.

The experimentally observed optical conductivity has other particular
features that are not contemplated by our two-band based model, such
as high-frequency interband transitions, a seemingly doping-independent
incoherent contribution with a rather long tail, other possible low-weight
Drude-like peaks at finite frequencies\cite{Uchida10,Lucarelli10}
and, of course, the origin of the scattering processes that lead to
a finite lifetime $\tau$. Clearly, a detailed description of the
optical spectrum has to take into account the effects of the other
bands that do not participate in the spin density wave state\cite{Kaneshita09}, and
possibly the role played by different orbitals that cross the Fermi
level. Yet, our simplified model that provides a very satisfactory
description\cite{Fernandes10} of the phase diagram of $\mathrm{Ba}(\mathrm{Fe}_{1-x}\mathrm{Co}_{x})_{2}\mathrm{As}_{2}$,
is able to correctly capture not only the main qualitative features
of the spectrum, but also the order of magnitude of the frequency
associated to the MIR peak.

Most interestingly, our results clarify how spectral weight is transferred
in magnetic superconductors below the superconducting transition temperature,
even in case where the Fermi surface of the ordered magnet is fully
gapped. In classical superconductors, only the spectral weight below
the optical gap $E_{g}=2\Delta$ is transferred to the Drude peak.
However, in the case where itinerant magnetism is also present and
particle-hole symmetry holds, the optical gap is given by $E_{g}=2\sqrt{\Delta_{\mathrm{AF}}^{2}+\Delta^{2}}$,
involving energies that can potentially be much larger than the superconducting
gap. Thus, in the regime where magnetism and superconductivity coexist
in the iron arsenides, the remaining Drude peak of the antiferromagnetic
phase, whose plasma frequency can be significantly smaller than the
plasma frequency of the paramagnetic state, is not the only origin
for the value of the superfluid condensate. Instead, spectral weight
associated to the higher-frequency MIR peak is transferred to the
$\delta$-function at $\omega=0$, enhancing the superfluid condensate.
Yet, this superfluid density is always smaller than its value in the
non-magnetic superconducting phase, in agreement with experiments\cite{Gordon10}.

This transfer of optical spectral weight is a consequence of the unique
rigidity of the superconductor with respect to transverse current
fluctuations. It implies that, even in the limit where the pure $T=0$
antiferromagnetic phase has no Drude peak, it is still possible to
obtain a finite superfluid density below $T_{c}$. The superfluid
density of the coexistence state is not only associated to electronic
states from the remaining Fermi surface, what allows the superconducting
transition to take place even when a Fermi surface would not be present
in the magnetically ordered state. This non-trivial observation is
corroborated by recent calculations of Vorontsov \emph{et al.}\cite{Vorontsov10},
that found coexisting itinerant magnetism and $s^{+-}$ superconductivity
in cases where the reconstructed Fermi surface would be completely
gapped at $T=0$. As they pointed out, in these situations both the
antiferromagnetic and superconducting phases are {}``effectively
attractive'' and cooperate to form the coexistence state.

\bigskip{}

We are grateful to R. Gordon and R.Prozorov for helpful discussions
and for sharing their penetration depth data prior to publication.
This research was supported by the Ames Laboratory, operated for the
U.S. Department of Energy by Iowa State University under Contract
No. DE-AC02-07CH11358.

\bigskip{}

\appendix

\section{Calculation of the penetration depth in the coexistence region}

Using Kramers-Kronig relations, the current-current correlation function
(\ref{current_current}) at finite momentum and finite frequency can
be written as:

\begin{widetext}

\begin{equation}
\Pi_{\alpha\beta}\left(\mathbf{q},i\omega_{n}\right)=e^{2}T\sum_{\mathbf{k},\nu_{m}}\int_{-\infty}^{\infty}\frac{d\nu}{\pi}\int_{-\infty}^{\infty}\frac{d\nu'}{\pi}\mathrm{tr}\left(\widehat{v}_{\mathbf{k+q}\alpha}\frac{\mathrm{Im}\widehat{G}_{\mathbf{k}+\mathbf{q}}\left(\nu'+i0^{+}\right)}{i\nu_{m}+i\omega_{n}-\nu'}\widehat{v}_{\mathbf{k}\beta}\frac{\mathrm{Im}\widehat{G}_{\mathbf{k}+\mathbf{q}}\left(\nu+i0^{+}\right)}{i\nu_{m}-\nu}\right)\label{general_pi}\end{equation}

It is straightforward to evaluate the Matsubara sum. Setting $\omega_{n}=0$
yields:

\begin{equation}
\Pi_{\alpha\beta}\left(\mathbf{q},\omega=0\right)=e^{2}\sum_{\mathbf{k}}\int_{-\infty}^{\infty}\frac{d\nu}{\pi}\int_{-\infty}^{\infty}\frac{d\nu'}{\pi}\left[\frac{n_{F}\left(\nu\right)-n_{F}\left(\nu'\right)}{\nu-\nu'}\right]\mathrm{tr}\left(\widehat{v}_{\mathbf{k+q}\alpha}\mathrm{Im}\widehat{G}_{\mathbf{k}+\mathbf{q}}\left(\nu'+i0^{+}\right)\widehat{v}_{\mathbf{k}\beta}\mathrm{Im}\widehat{G}_{\mathbf{k}+\mathbf{q}}\left(\nu+i0^{+}\right)\right)\label{aux_general_pi}\end{equation}
 where $n_{F}$ is the Fermi function. The imaginary part of the Green's
function can be calculated directly from Eq. (\ref{greens_function_phs}):

\begin{equation}
\widehat{G}_{\mathbf{k}}\left(\nu+i0^{+}\right)=\frac{\nu\tau_{0}\sigma_{0}+\widehat{\varepsilon}_{\mathbf{k}}}{2E_{\mathbf{k}}}\left(\frac{1}{\nu-E_{\mathbf{k}}+i0^{+}}-\frac{1}{\nu+E_{\mathbf{k}}+i0^{+}}\right)\label{aux_greens_function}\end{equation}
 where we defined the positive excitation energy $E_{\mathbf{k}}=\sqrt{\xi_{\mathbf{k}}^{2}+\Delta_{\mathrm{AF}}^{2}+\Delta^{2}}$.
It follows that:

\begin{equation}
\mathrm{Im}\widehat{G}_{\mathbf{k}}\left(\nu+i0^{+}\right)=-\frac{\pi\left(\nu\tau_{0}\sigma_{0}+\widehat{\varepsilon}_{\mathbf{k}}\right)}{2E_{\mathbf{k}}}\left[\delta\left(\omega-E_{\mathbf{k}}\right)-\delta\left(\omega+E_{\mathbf{k}}\right)\right]\label{imaginary_greens_function}\end{equation}

Substituting Eq. (\ref{imaginary_greens_function}) in the expression
(\ref{aux_general_pi}), we can evaluate the frequency integrals as
well as the trace in Nambu space. In the limit of small momentum,
we obtain:

\begin{eqnarray}
\Pi_{\alpha\beta}\left(\mathbf{q}\rightarrow0,\omega=0\right) & = & \lim_{\mathbf{q}\rightarrow0}\sum_{\mathbf{k}}2e^{2}v_{\mathbf{k}}^{\alpha}v_{\mathbf{k}}^{\beta}\left\{ \left[\frac{n_{F}\left(E_{\mathbf{k}}\right)-n_{F}\left(E_{\mathbf{k}+\mathbf{q}}\right)}{E_{\mathbf{k}}-E_{\mathbf{k}+\mathbf{q}}}\right]\left(1+\frac{\xi_{\mathbf{k}}\xi_{\mathbf{k}+\mathbf{q}}+\Delta^{2}-\Delta_{\mathrm{AF}}^{2}}{E_{\mathbf{k}}E_{\mathbf{k}+\mathbf{q}}}\right)\right.\nonumber \\
 &  & \left.+\left[\frac{n_{F}\left(E_{\mathbf{k}}\right)-n_{F}\left(-E_{\mathbf{k}+\mathbf{q}}\right)}{E_{\mathbf{k}}+E_{\mathbf{k}+\mathbf{q}}}\right]\left(1-\frac{\xi_{\mathbf{k}}\xi_{\mathbf{k}+\mathbf{q}}+\Delta^{2}-\Delta_{\mathrm{AF}}^{2}}{E_{\mathbf{k}}E_{\mathbf{k}+\mathbf{q}}}\right)\right\} \label{aux_Pi_final}\end{eqnarray}
 yielding, for a two-dimensional isotropic superconductor:

\begin{eqnarray}
\Pi_{\alpha\beta}\left(\mathbf{q}\rightarrow0,\omega=0\right) & = & -v_{F}^{2}e^{2}\delta_{\alpha\beta}\sum_{\mathbf{k}}\left[2\left(-\frac{\partial n_{F}}{\partial E_{\mathbf{k}}}\right)\left(1-\frac{\Delta_{\mathrm{AF}}^{2}}{E_{\mathbf{k}}^{2}}\right)+\frac{\Delta_{\mathrm{AF}}^{2}}{E_{\mathbf{k}}^{3}}\tanh\left(\frac{\beta E_{\mathbf{k}}}{2}\right)\right]\label{aux_Pi_final_2}\end{eqnarray}

We introduce the density of states $\rho$ and take $T=0$, obtaining:

\begin{equation}
\Pi_{\alpha\beta}\left(\mathbf{q}\rightarrow0,\omega=0\right)=-\rho v_{F}^{2}e^{2}\delta_{\alpha\beta}\int_{-\infty}^{\infty}d\xi\left[2\delta\left(\sqrt{\xi^{2}+\Delta^{2}+\Delta_{\mathrm{AF}}^{2}}\right)\left(1-\frac{\Delta_{\mathrm{AF}}^{2}}{\xi^{2}+\Delta^{2}+\Delta_{\mathrm{AF}}^{2}}\right)+\frac{\Delta_{\mathrm{AF}}^{2}}{\left(\xi^{2}+\Delta^{2}+\Delta_{\mathrm{AF}}^{2}\right)^{3/2}}\right]\label{Pi_final}\end{equation}

\end{widetext}

For $\Delta\neq0$ or $\Delta_{\mathrm{AF}}\neq0$ the first term
vanishes, whereas the second one gives:

\begin{equation}
\Pi_{\alpha\beta}\left(\mathbf{q}\rightarrow0,\omega=0\right)=-2\rho v_{F}^{2}e^{2}\delta_{\alpha\beta}\left(\frac{\Delta_{\mathrm{AF}}^{2}}{\Delta^{2}+\Delta_{\mathrm{AF}}^{2}}\right)\label{pi_T_0}\end{equation}
 which leads to the same result as Eq. (\ref{pen res}) from the main
text. Notice, from Eq. (\ref{aux_Pi_final}), that the non-zero value
assumed by the current-current correlation function at $T=0$ is due
to the coherence factor $\left(1-\frac{\xi_{\mathbf{k}}\xi_{\mathbf{k}+\mathbf{q}}+\Delta^{2}-\Delta_{\mathrm{AF}}^{2}}{E_{\mathbf{k}}E_{\mathbf{k}+\mathbf{q}}}\right)$.
For a non-magnetic superconductor, this term goes to zero as $\mathbf{q}\rightarrow0$
for any temperature; then, the only contribution to the current-current
correlation function comes from the usual coherence factor $\left(1+\frac{\xi_{\mathbf{k}}\xi_{\mathbf{k}+\mathbf{q}}+\Delta^{2}-\Delta_{\mathrm{AF}}^{2}}{E_{\mathbf{k}}E_{\mathbf{k}+\mathbf{q}}}\right)$,
whose prefactor vanishes at $T=0$ due to the existence of a gap in
the quasiparticle energy spectrum. In both coherence factors, the
relative minus sign between $\Delta^{2}$ and $\Delta_{\mathrm{AF}}^{2}$
is a result of the fact that while $\xi$ and $\Delta$ change from
one Fermi surface sheet to the other, $\Delta_{\mathrm{AF}}$ stays
the same. Therefore, the change in the penetration depth cannot be
attributed to a change only in the density of states, in accordance
to our analysis of the finite frequency optical spectrum.

\end{document}